\documentclass[floatfix,12pt]{iopart}
\usepackage{iopams}
\usepackage{graphicx}
\usepackage{subfigure}

\begin{document}

\title{Random walkers with extreme value memory: modelling the peak-end rule}
\author{Rosemary J\ Harris}
\address{School of Mathematical Sciences, Queen Mary University of London, Mile End Road, London, E1 4NS, UK}
\ead{\mailto{rosemary.harris@qmul.ac.uk}}

\begin{abstract}
Motivated by the psychological literature on the ``peak-end rule''
for remembered experience, we perform an analysis within a random walk
framework of a discrete choice model where agents' future choices depend on
the peak memory of their past experiences. In particular, we use this
approach to investigate whether increased noise/disruption always leads to
more switching between decisions.  Here extreme value theory illuminates
different classes of dynamics indicating that the long-time behaviour is
dependent on the scale used for reflection; this could have implications,
for example, in questionnaire design.

\end{abstract}

\section{Introduction}

The use of stochastic processes in interdisciplinary modelling has a long history dating back at least to Bachelier's seminal work in finance~\cite{Bachelier1900} and encompassing applications to traffic flow~\cite{Chowdhury00}, biological processes~\cite{Bressloff14} and opinion dynamics~\cite{Castellano09b}, among others.  Often such systems are analysed using a Markovian (or memoryless) approximation which considerably simplifies the theoretical treatment.  However, within the statistical mechanics community there is much topical interest in characterizing the properties of non-Markovian models.   There are many ways to incorporate memory effects including generalized Langevin or Fokker-Planck approaches~\cite{Hanggi78,Volkov83,Metzler99}, and the assumption of internal variables or non-exponential waiting times in many-particle microscopic models~\cite{Hirschberg09,Concannon14,Khoromskaia14}.  At the random walk level, recent analytical studies in the physics literature have included the imaginatively named ``elephant'' random walker who remembers a property of the entire history~\cite{Schutz04},  the ``Alzheimer'' random walker who recalls just the distant past~\cite{Cressoni07,Kenkre07}, and ``bold'' and ``timorous'' random walkers who behave differently only when they are at the furthest point ever attained~\cite{Serva13b}.  In fact, the elephant random walk can also be related to the older P\'olya urn problem~\cite{Polya31}; see~\cite{Pemantle07} for a mathematical review of this and other random processes with reinforcement.  Amongst the more recent rigorous results are some for the ``excited'' (or ``cookie'') random walk~\cite{Benjamini03,Basdevant08}, random walks with different kinds of self-interaction~\cite{Veto08,Erschler12}, and those with internal states~\cite{Nandori11}.

In real-life social and economic scenarios the dependence on memory is, of course, rather complex.  However, one psychological heuristic is the ``peak-end rule'' suggested by Kahneman et al.~\cite{Kahneman93}.  This asserts that the remembered utility (loosely speaking the pleasure or pain experienced) of a specific situation/episode is approximately given by the average of the peak experience (best or worst) during the event and the final experience of that event.  Notice in particular that this implies ``duration neglect''~\cite{Fredrickson08} in the sense that the extreme and final snapshots are considerably more important in the memory than the overall length of the experience (even if it is an unpleasant one!).  Empirical support for this peak-end approximation comes from situations ranging from the pain of medical procedures~\cite{Redelmeier96} to the pleasure of material goods~\cite{Do08}.  Whilst other work paints a more complicated picture, particularly for extended events~\cite{Kemp08}, it is clear that peak experiences play an important role and, to the best of our knowledge, such memory of extreme values is largely unexplored from the perspective of statistical physics.\footnote{Note that this is subtlely different from problems in which the dynamics depends on prediction of \emph{future} extreme values as relevant, for example, in financial applications.}

In this spirit, our contribution is to consider a random walk model where the probability of moving left or right depends on the maximum value of a random variable associated to each time step.  As we will show, this can be thought of as a simple discrete choice model with a dependence on the ``peak'' of past experience.  In particular, we use this framework to investigate whether increased noise in the model (corresponding perhaps to the ``churn'' of changing circumstances or some kind of disruption, cf.\ e.g.,~\cite{Marsden13}), always leads to more switching between decisions.  Using the mathematics of extreme values, we show that the answer to this question depends on the distribution of the random variable encoding the experience at each step.  Our work thus helps to shed light on real-world issues as well as contributing to building up general understanding of memory effects in statistical mechanics models.

The remainder of the paper is structured as follows.  In section~\ref{s:setup} we describe our random walk formalism and explain its significance as an opinion choice model as well as the manner in which it extends previous work on generalized P\'olya urns.  In section~\ref{s:extreme} we employ extreme value theory to develop a heuristic argument for different classes of long-time behaviour depending on the distribution of past experience, and compare our predictions with simulations.   Finally, in section~\ref{s:dis}, we conclude with a discussion of implications and open questions. 

\section{Random walk set-up and interpretation as decision model}
\label{s:setup}

We consider a one-dimensional random walker who steps right or left in discrete time, denoting by $X^+_t$ the number of steps right up to time $t$ and $X^-_t$ the corresponding number left.  Note that $X^-_t=t-X^+_t$ by construction.  For later convenience, we also define the corresponding time averages (``velocities'') $V^+\equiv X^+/t $ and $V^- \equiv X^- /t$, suppressing the notational dependence on $t$ where no confusion should arise.   

In addition, at each time step $i$ we associate an independent identically distributed (i.i.d.) random variable $U_i$ from some known distribution with cumulative distribution function (c.d.f.) $F(u)$.  Crucially, the walker ``remembers'' the maximum value of $U_i$ for all rightward steps in its history and, separately, the maximum value of $U_i$ for all leftward steps.  We denote these history-dependent random variables by $U^+_t$ and $U^-_t$ respectively so that formally we have 
\begin{equation}
U^\pm_t = \max_{1 \leq i\leq t}\{U_i : X^\pm_i - X^\pm_{i-1} = 1\}.
\end{equation}
Memory is then built into the dynamics via the setting of left and right hopping probabilities for the next step to depend on the current values of $U^+$ and $U^-$.  It is clear that the system is non-Markovian in position space although, of course, still Markovian in an enlarged state space including $U^+$ and $U^-$.

The central idea is that this set-up is analogous to a single agent in a discrete decision model where $U$ is some kind of ``utility'' and the agent remembers its extreme value (corresponding to the ``peak'' part of Kahneman's peak-end rule) for each of two choices.  Specifically, we fix the right and left stepping probabilities as functions of the random variables $U^+$ and $U^-$  to accord with the familiar ``logit'' choices of economic theory
\begin{equation}
P^+ = \frac{\rme^{U^+/T}}{\rme^{U^+/T}+\rme^{U^-/T}}, \qquad P^- = \frac{\rme^{U^-/T}}{\rme^{U^+/T}+\rme^{U^-/T}} \label{e:logit} 
\end{equation}
where the positive parameter $T$ represents the level of noise in the decision.\footnote{The symbol $T$ is chosen here to reinforce the analogy with temperature in a physical system and is not to be confused with a time parameter; the update probabilities~\eref{e:logit} loosely resemble the Glauber refreshment formulae for an Ising model~\cite{Glauber63}.}  Throughout the paper we set $U^+_0=U^-_0=0$ so that the two choices (step directions) are initially equally likely; as the system evolves the jump probabilities become asymmetric due to differing values of $U^+$, and $U^-$.  In particular, note that $U^+$ and $U^-$ are monotonically increasing with the number of steps right and left respectively. 

In passing, we note here that if $U^\pm$ were deterministic functions of the velocities $V^\pm$ the model would closely resemble the P\'olya urn problem, familiar in the mathematics literature~\cite{Polya31}, where the probability of selecting a ball of a particular colour depends on the fraction of that colour chosen previously (in a similar manner, the elephant random walker of~\cite{Schutz04} steps left and right with probabilities depending on the relative number of such steps in his past).  If, as here, the probability function is nonlinear, the urn model is known as a generalized P\'olya process~\cite{Pemantle07,Hill80,Mori15}.  The crucial difference in our model is that $U^\pm$ fluctuate in a correlated way due to the statistics of the extreme values -- we seek to determine the effect this has on long-time properties such as the average velocity of the random walker (or, equivalently, the proportion of time the agent makes each decision).

One might naively expect that, as for the P\'olya models, in the large $t$ limit our random walker approaches a fixed-point state where the relative probabilities (and hence the fraction of steps left and right) do not change.  The symmetry of~\eref{e:logit} suggests two specific types of fixed point: (i) $(V^+,V^-)=(1/2,1/2)$ with $U^+$ and $U^-$ asymptotically equal and hence symmetric behaviour of the random walker, i.e., both choices equally likely in the long run; (ii) $(V^+,V^-)=(0,1)$ or $(V^+,V^-)=(1,0)$ with one of $U^+$ or $U^-$ negligibly small with respect to the other and hence an asymmetric random walker moving only in one direction, i.e., the agent frozen in one or other choice.

We shall demonstrate the existence of these fixed points more carefully later.  For now, we remark that a pertinent question relates to their stability, in particular, whether the symmetric fixed point $(1/2,1/2)$ can be made stable by increasing the noise.  This could be important for sharing the load between two different choices (e.g., two different routes or transport options).  In the next section, we address this issue for different distributions of $U$ before,  in section~\ref{s:dis}, considering the added effect of the ``end'' part of the peak-end rule.

\section{Extreme value controlled behaviour}
\label{s:extreme}

\subsection{Outline of method}
\label{ss:method}

The behaviour of the model will obviously depend on the distribution of $U$.   Our strategy is first to analyse the typical long-time dynamics by approximating $U^\pm$ in~\eref{e:logit} by the so-called ``characteristic largest value'' of extreme value theory and then, where relevant, to consider the added effect of fluctuations about this.  

The characteristic largest value after $X^\pm$ trials is defined for a given $F(u)$ as the value of $u$ at which $F(u) = 1 - 1/X^\pm$.  It gives a straightforward way to obtain the scaling of the maximum value and is closely related to other properties of the full distribution~\cite{Gumbel04}, as we shall see for various cases in the following subsections.  Our approach using the characteristic largest value leads to hopping probabilities depending on the number of left/right steps over the whole previous history and is thus in the spirit of the generalized P\'olya urn models mentioned above or continuous-time analogues with current-dependent hopping rates~\cite{Me09}.  One subtlety here is that the resulting probabilities in our model depend directly on the number of steps left and right, $X^\pm$, not the fractions, $V^\pm$.  Depending on the functional form of the characteristic largest value this may introduce an explicit time dependence in the dynamics for $V^\pm$ as we shall see in some of the subsequent examples.

\begin{figure}
\centering
\subfigure[\ $\Delta'(v^*)<1$]{\includegraphics[width=0.4\textwidth]{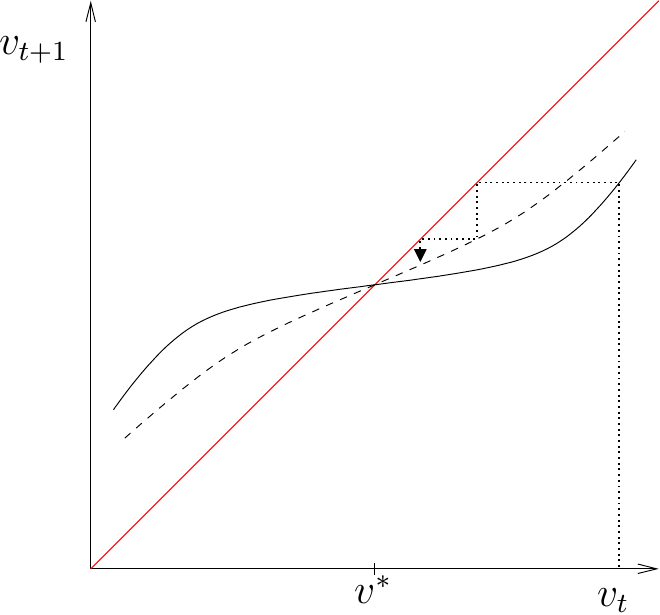}} \hfill
\subfigure[\ $\Delta'(v^*)>1$]{\includegraphics[width=0.4\textwidth]{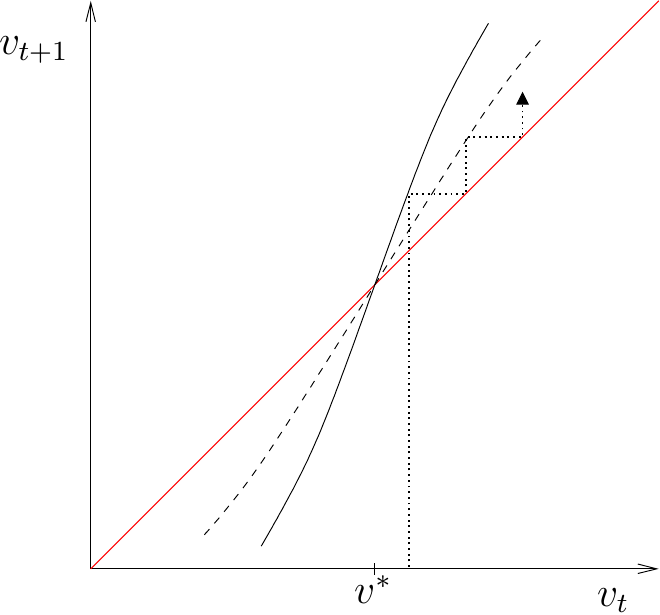}}
\caption{Sketch graphs showing iteration of $v_t$ using~\eref{e:dismap} in vicinity of stable fixed point (left) and unstable fixed point (right).  The slope of $v_{t+1}$ as a function of $v_t$ is $(t+\Delta'(v_t))/(t+1)$ which approaches unity as $t$ increases (indicated by dashed lines) leading to slow decay/growth.}
\label{f:cobweb}
\end{figure}

In fact, since $V^++V^-=1$ by construction, this procedure enables us for a given utility distribution $F(u)$ to write $P^\pm$ simply in terms of $V\equiv V^+-V^-$ and possibly time $t$.  Now, if the random variable $V$ takes value $v$, the mean distance moved in the next step is given by the corresponding value of $P^+-P^-$ which we denote by $\Delta_t(v)$.   Hence, on average, we expect a ``typical trajectory'' given by the discrete mapping  
\begin{equation}
v_{t+1}=\frac{t v_t + \Delta_t(v_t)}{t+1}. \label{e:dismap}
\end{equation}
For cases where $\Delta_t(v)$ has no dependence on $t$ it is immediately clear from~\eref{e:dismap} that fixed points $v^*$ should satisfy $v^* = \Delta(v^*)$ and a standard ``cobweb''-type construction predicts that the stability of the fixed points is determined by the slope of the increasing function $\Delta(v)$, see figure~\ref{f:cobweb}.    
Specifically, if $\Delta'(v^*)<1$ [figure~\ref{f:cobweb}(a)] then small fluctuations below $v^*$ are characterized by $\Delta(v)>v$, so on average the velocity increases back towards the fixed point.  Similarly, fluctuations above $v^*$ have $\Delta(v)<v$, so on average the velocity decreases, again back towards the fixed point.  An analogous argument shows that fixed points with $\Delta'(v^*)>1$ are unstable [figure~\ref{f:cobweb}(b)].  Notice that due to the time dependence of the mapping~\eref{e:dismap} the decay towards fixed points is expected to be power-law rather than exponential in nature -- physically this is because as the measurement time increases the last step has a smaller and smaller effect on the overall time average. 

In the following subsections we illustrate this approach for three qualitatively different scenarios corresponding to the three known families of extreme value theory.  (In cases where $\Delta_t(v)$ itself depends on time we shall chiefly be interested in its behaviour as $t \to \infty$.)  We then confront the predictions with simulation results and discuss how fluctuations in the extreme values modify the picture of typical behaviour given above.

\subsection{Exponential tails}
\label{ss:exp}

To demonstrate the method, we first look in detail at the case where the utility variable $U$ has an exponential distribution with c.d.f.\
\begin{equation}
F(u)=1 - \rme^{-\lambda u}, \quad u\geq 0. \label{e:exp}
\end{equation}
Here the characteristic largest value after $X^\pm$ steps is given by $({\ln X^\pm})/{\lambda}$ so, substituting for $U^\pm$ in~\eref{e:logit}, we approximate $P^\pm$ in the long-time limit by
\begin{equation}
P^\pm = \frac{(X^\pm)^{1 /(\lambda T)}}{(X^+)^{1 /(\lambda T)}+(X^-)^{1 /(\lambda  T)}}
\end{equation}
or equivalently, in terms of the time averages, 
\begin{equation}
P^\pm = \frac{(V^\pm)^{1 /(\lambda T)}}{(V^+)^{1 /(\lambda T)}+(V^-)^{1 /(\lambda T)}}.
\end{equation}
In this particular case, the probabilities can be written in terms of $V^\pm$ without explicit time dependence illustrating a direct connection to the class of elephant random walker and (time-homogeneous) P\'olya urn problems.

To determine the fixed points we further write $P^\pm$ in terms of the net velocity by substituting $V^\pm=(1\pm V)/2$ to obtain
\begin{equation}
P^+-P^-=\frac{(1+V)^{ 1 /(\lambda  T)} - (1-V)^{1 /(\lambda T)}}{(1+V)^{1 /(\lambda T)} + (1-V)^{1 /(\lambda T)}},
\end{equation}
so the function specifying the mean displacement of the next step can be compactly written as
\begin{equation}
\Delta(v) = \tanh\left( \frac{1 }{2 \lambda T} \ln \frac{1+v}{1-v} \right) \label{e:Delta}
\end{equation}
where $v$ is the current value of the velocity.  The fixed points satisfying $v^* = \Delta(v^*)$ are then seen by inspection to be $v^* = 0, \pm 1$ as predicted from symmetry arguments. 

\begin{figure}
\centering
\includegraphics[width=0.8\textwidth]{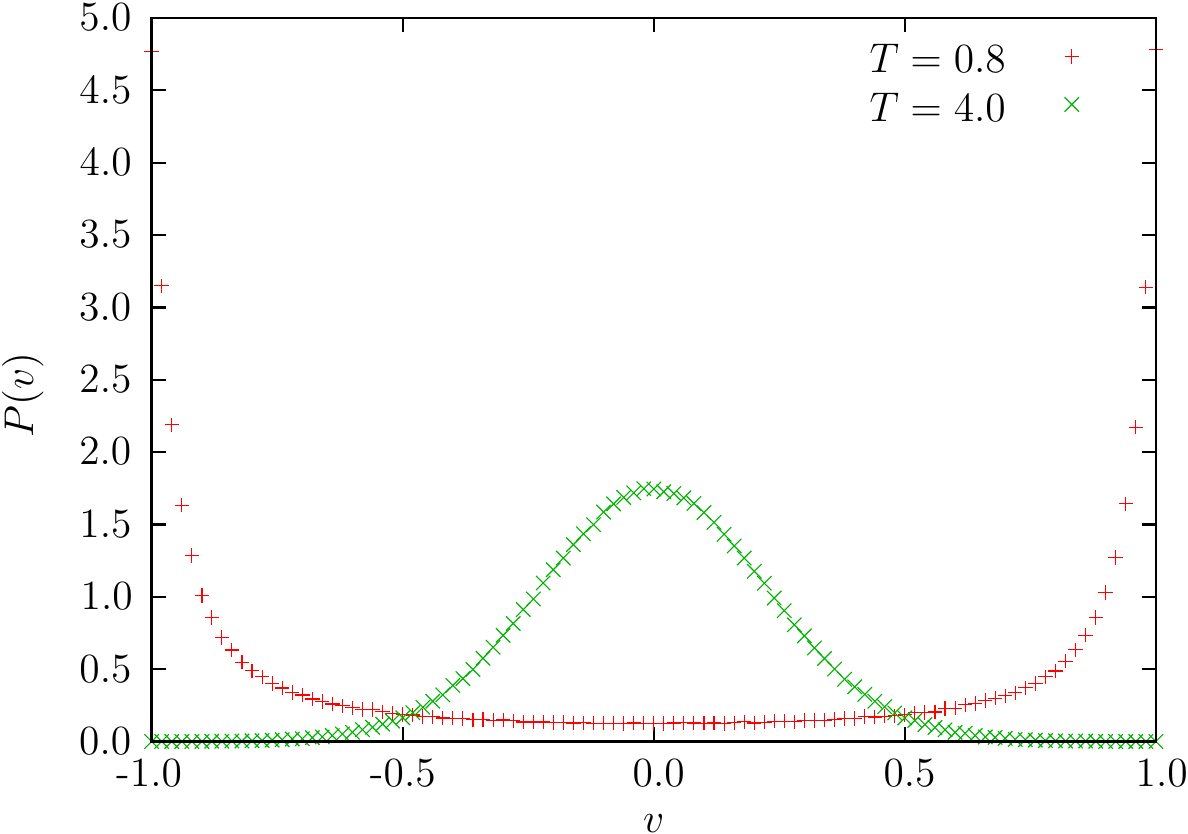}
\caption{Histogram of final velocity $V=V^+-V^-$ for random walkers with dependence on peak values of exponential utility distribution ($\lambda=1$) and two different noise levels ($T=0.8$ and $T=4.0$).  Distribution calculated from $10^6$ trajectories each running up to final time $t=100$.}
\label{f:expeg}
\end{figure}

Recall from the previous subsection that to determine which of these fixed points is stable we need to check the slope $\Delta'(v^*)$; here it is straightforward to show from~\eref{e:Delta} that
\begin{equation}
\Delta'(0) = \frac{1}{\lambda T}.
\end{equation}
Hence if $\lambda T > 1 $, the mixed solution $(V^+,V^-)=(1/2,1/2)$ is stable and the asymmetric frozen solutions $(V^+,V^-)=(1,0)$ and $(V^+,V^-)=(0,1)$ correspondingly unstable.  Similarly, for $\lambda T < 1$, the mixed solution is unstable and we predict that the random walker becomes frozen into ballistic motion in one of the two directions.   To check this heuristic argument we appeal to Monte Carlo simulations -- in figure~\ref{f:expeg} we show the empirical distribution of velocities at $t=100$ for an exponential utility distribution (with mean $\lambda=1$) and values of noise predicted to correspond to the two different cases ($T=0.8$ and $T=4.0$).  

\begin{figure}
\centering
\includegraphics[width=0.8\textwidth]{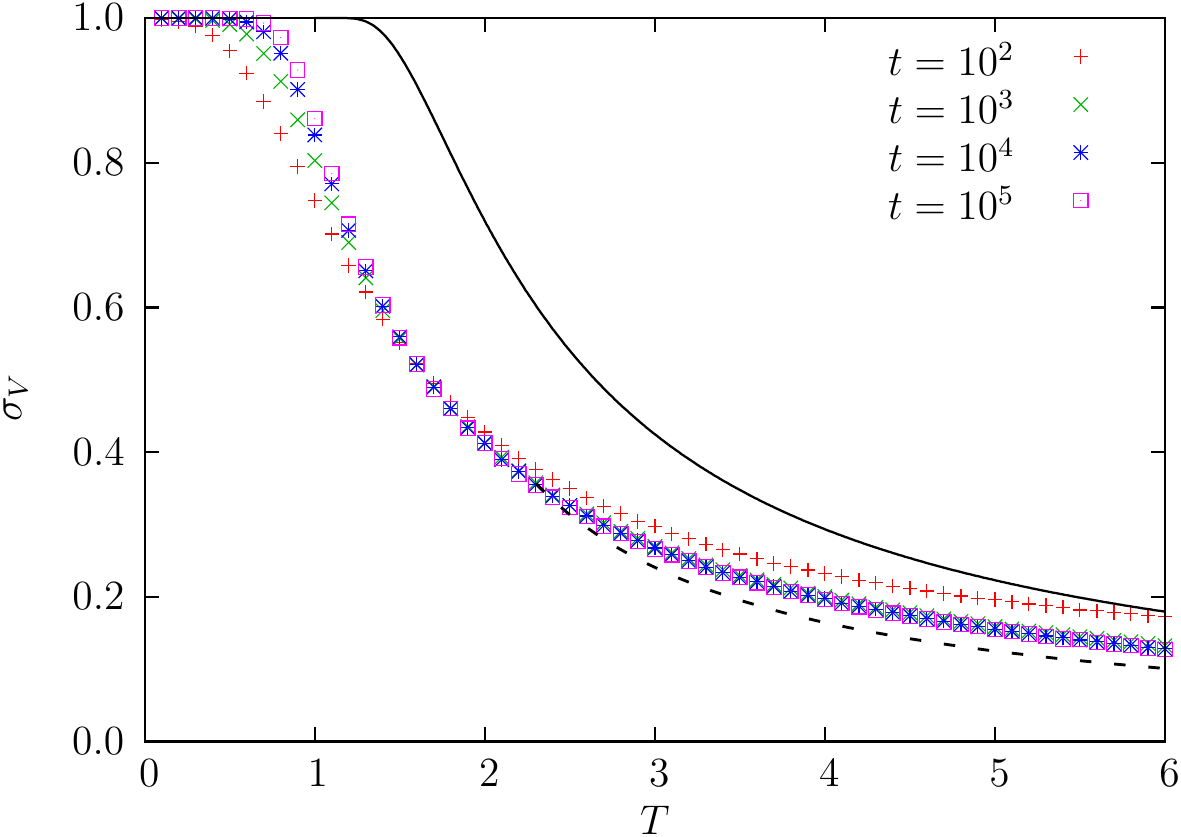}
\caption{Standard deviation of final velocity $V=V^+-V^-$ for random walkers with dependence on peak values of exponential utility distribution ($\lambda=1$) and range of noise values.  Points show simulation results for increasing times ($10^5$ trajectories in each case); solid line is numerical solution of~\eref{e:crude}; dashed line is approximation~\eref{e:finapprox}.}
\label{f:range}
\end{figure}

We see good qualitative agreement of the simulations with the prediction: in the low-noise case the trajectories are sharply peaked around the asymmetric fixed points, i.e., $V=V^+-V^-=\pm 1$ (corresponding to each agent almost always making the same choice) whilst in the high-noise case the trajectories are clustered around the symmetric fixed point, i.e., $V=V^+-V^-=0$ (corresponding to each agent sampling the two choices approximately equally).  However, there is a finite width of the distribution about the fixed point(s) even for $\lambda T$ significantly greater than unity -- to investigate this more systematically, and reveal possible finite-time effects, we plot in figure~\ref{f:range} the standard deviation of the distribution as a function of $T$ for increasing measurement times.   
This quantifies how close the trajectories end to symmetric or asymmetric fixed points without making a distinction between the two asymmetric states (whose selection is expected to depend sensitively on the agent's first few choices).  According to the analysis of typical behaviour given above, we expect the standard deviation to be unity for $\lambda T < 1$ and zero for $\lambda T > 1$.  In fact, although the simulations do show evidence of a transition around $\lambda T = 1$, the situation is somewhat more complicated; in particular, the standard deviation clearly converges to a finite value even for $\lambda T > 1$.

These observed results for the standard deviation $\sigma_V$ suggest that, even in the long-time limit, the properties of the model are sensitive to the full distribution of maximum values not just the characteristic largest value.  As further evidence of this, we remark that if $U^\pm$ \emph{were} given deterministically by the characteristic largest value, the variance for $\lambda T > 1$ could only be due to decay towards the stable fixed point and fluctuation of individual trajectories about the typical behaviour.   In this case, $V$ would be expected to obey a large deviation principle with some ``speed'' $t^\alpha$~\cite{Me09,Touchette09b} and the variance would eventually converge to zero, as confirmed in~\ref{A:det} where, for comparative purposes, we present simulation results from an artificial model with $U^\pm$ at every time step set equal to $({\ln X^\pm})/{\lambda}$.  It is clear then that the limiting value of the variance in the full model is determined by fluctuations in the extreme values, leading to fluctuations of the typical trajectories themselves. 

In the case of an exponential distribution it is, of course, well known that the limiting form of the rescaled maximum has a Gumbel distribution; here the c.d.f.\ of $U^\pm$ is asymptotically given by  
\begin{equation}
G(u^\pm)=\rme^{-\rme^{-(u^\pm-a^\pm)/b^\pm}}
\end{equation}
where $a^\pm={\ln X^\pm}/{\lambda}$ and $b^\pm=1/\lambda$ (see, e.g.,~\cite{Gumbel04,deHaan06} and references therein).  The mode $a^\pm$ coincides with the characteristic largest value calculated earlier while the mean is $a^\pm+b^\pm\gamma$ (with $\gamma$ the Euler-Mascheroni constant) so differs from it only by a constant amount.  Taking account of the fluctuations, the maximum value random variables thus obey
\begin{equation}
U^+-U^-=\frac{\ln X^+}{\lambda} - \frac{\ln X^-}{\lambda} + \epsilon \label{e:eps}
\end{equation}
where the distribution of $\epsilon$ is given by the difference of the two Gumbel distributions as a logistic distribution with mean zero and variance $\sigma^2_\epsilon=\pi^2/(3 \lambda^2)$.  

\begin{figure}
\centering
\includegraphics[width=0.8\textwidth]{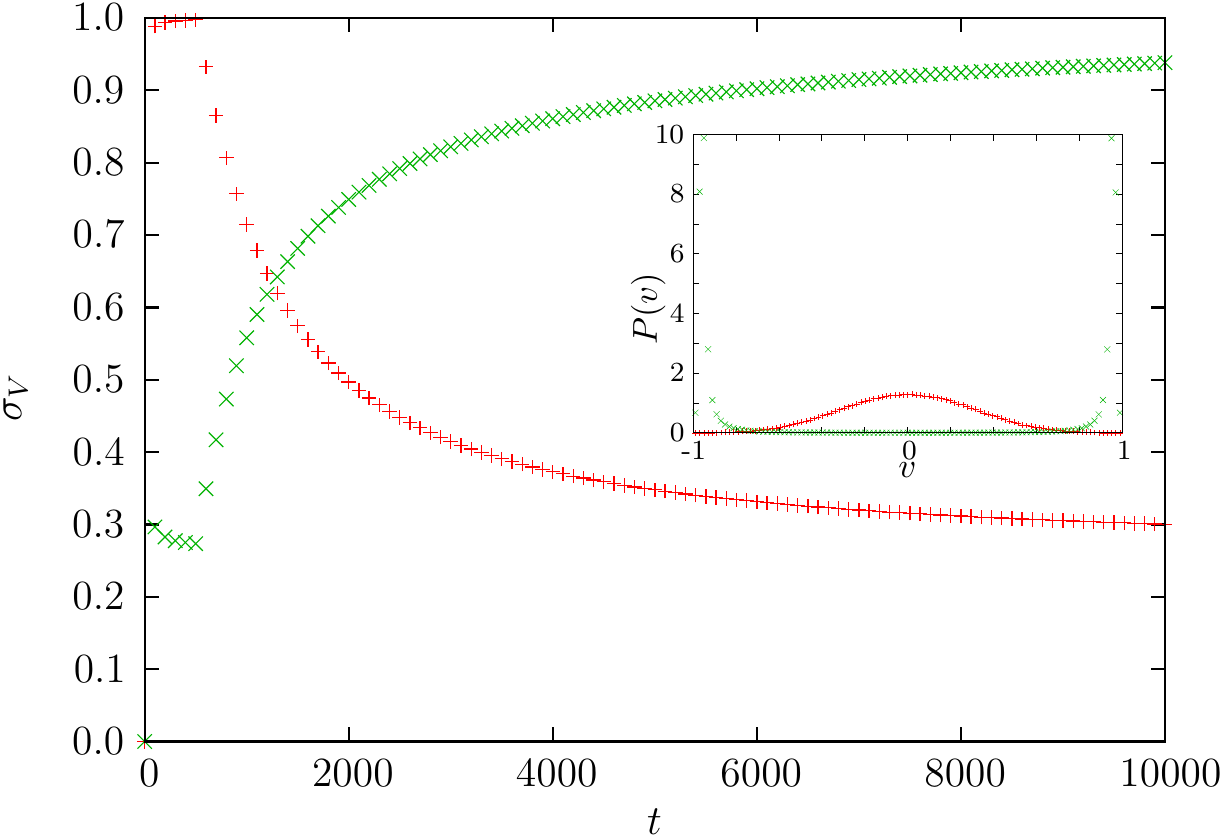}
\caption{Standard deviation of velocity $V=V^+-V^-$ against time for random walkers with dependence on peak values of exponential utility distribution ($\lambda=1$) and change in noise level at $t=500$: $T=0.3$ to $T=3.0$ in red ($+$ symbols), $T=3.0$ to $T=0.3$ in green  ($\times$ symbols).  Inset shows final ($t=10000$) velocity histograms in the two cases.  All calculations from $10^6$ trajectories.}
\label{f:control}
\end{figure}

Substituting the form of~\eref{e:eps} in the expression for $P^+-P^-$ and repeating the calculations leading to~\eref{e:Delta} one finds that for a given, non-zero, value of $\epsilon$ the position of the ``symmetric'' fixed point is shifted from zero although both its stability and the position of the asymmetric fixed points remain unchanged.  A crude estimate of the standard deviation in the position of the symmetric fixed point can be obtained as the value of $v \in (-1,1)$ which solves the transcendental equation
\begin{equation}
v=\tanh\left( \frac{1 }{2 \lambda T} \ln \frac{1+v}{1-v} + \frac{\sigma_\epsilon}{2T}  \right). \label{e:crude}
\end{equation}
As seen in figure~\ref{f:range} this leads to a loose upper bound on the observed standard deviation.  The actual standard deviation is smaller because the value of $\epsilon$ and the corresponding fixed point changes during the course of each trajectory.  In~\ref{A:var}, we include this effect within a linear expansion to obtain an analytical expression for $\sigma_V$ which is a better approximation for $\lambda T $ large (see, again, figure~\ref{f:range}).  Notwithstanding the finite variance, the claim that one can control the long-time behaviour by increasing/decreasing the noise is well borne out by simulation.  For example, in figure~\ref{f:control} we show the evolution of the standard deviation in a scenario where the noise level (and hence the stability of the fixed points) is abruptly changed after the first 500 time steps.

The cornerstone of extreme value theory, the Fisher-Tippett-Gnedenko theorem~\cite{Fisher28,Gnedenko43}, asserts that the Gumbel distribution is universal for the rescaled maximum of i.i.d.\ random variables drawn from a distribution with exponential tails.  However, the functional form of the scaling parameters depends on the distribution being considered.   As a second example, we now make the arguably reasonable hypothesis that agents assign utilities according to a Gaussian with some mean $\mu$ and standard deviation $\varsigma$.  The mode of the limiting distribution, again given by the characteristic largest value, is
\begin{equation}
a^\pm = \mu - \varsigma \Phi^{-1}({1}/{X^\pm})
\end{equation}
where $\Phi$ is the c.d.f.\ of the standard normal distribution.  We note that, in this case, $a^\pm$ retains a logarithmic dependence on $X^\pm$ growing like $\varsigma \sqrt{2 \ln X^\pm}$ as $X^\pm \to \infty$.  Ignoring the fluctuations about this value, an analogous argument to that given above then yields for large $t$  
\begin{equation}
\Delta_t(v) = \tanh\left( \frac{\varsigma }{2  T} \left\{ \sqrt{2\ln\left[\frac{(1+v)t}{2}\right]} -  \sqrt{2\ln\left[\frac{(1-v)t}{2}\right]} \right\} \right).
\end{equation}

\begin{figure}
\centering
\includegraphics[width=0.8\textwidth]{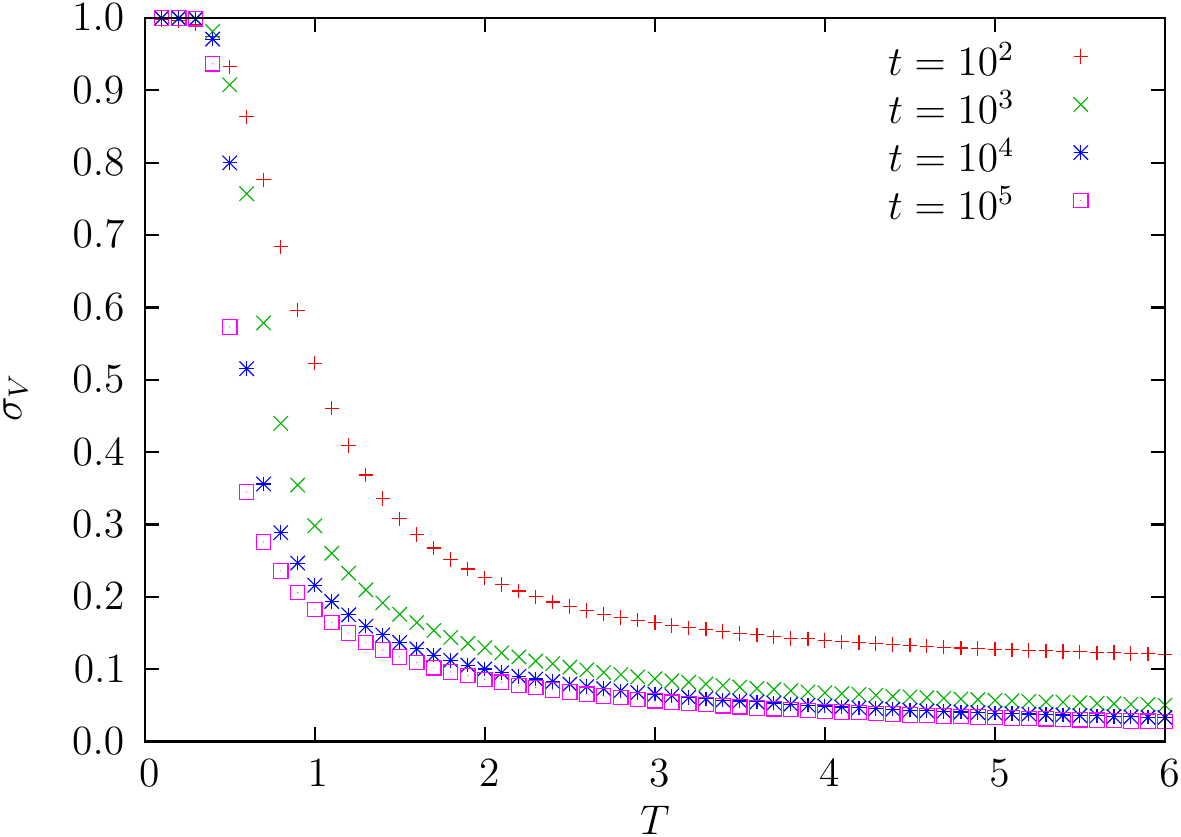}
\caption{
Same as figure~\ref{f:range} but for Gaussian utility distribution ($\mu=1$, $\varsigma=1$). 
}
\label{f:rangeg}
\end{figure}

It is clear that $v^*=0$ is a fixed point for all $t$ and its stability is controlled by the slope 
\begin{equation}
\Delta_t'(0) = \frac{\varsigma}{ T \sqrt{2 \ln (t/2)}}.
\end{equation}
As $t \to \infty$ the slope tends to zero and hence we predict that the symmetric fixed point is always stable in the long run.  However, since the dependence is only logarithmic in~$t$, one still expects to see a noise-controlled transition for large but finite times.  This is supported by the simulation results for standard deviation shown in figure~\ref{f:rangeg}.  For comparison, we have set there the first two moments equal to those of the exponential distribution in figure~\ref{f:range} and the picture for the Gaussian case is qualitatively similar with the transition between low and high-noise regimes only weakly dependent on $t$. 
To complete the story, we can again consider fluctuations of the maximum values.  In this case, the width of the Gumbel distribution is controlled by
\begin{equation}
b^\pm=-\frac{\varsigma}{\Phi^{-1}\left({1}/{X^\pm}\right)}
\end{equation}
which decays to zero as $X^\pm \to \infty$ (again see, e.g.,~\cite{Gumbel04,deHaan06}).  Hence, in contrast to the exponential case, we do not expect a finite limiting velocity variance in the high-noise regime and indeed the relevant simulation data do seem to show a slow convergence towards zero.

\begin{figure}
\centering
\includegraphics[width=0.8\textwidth]{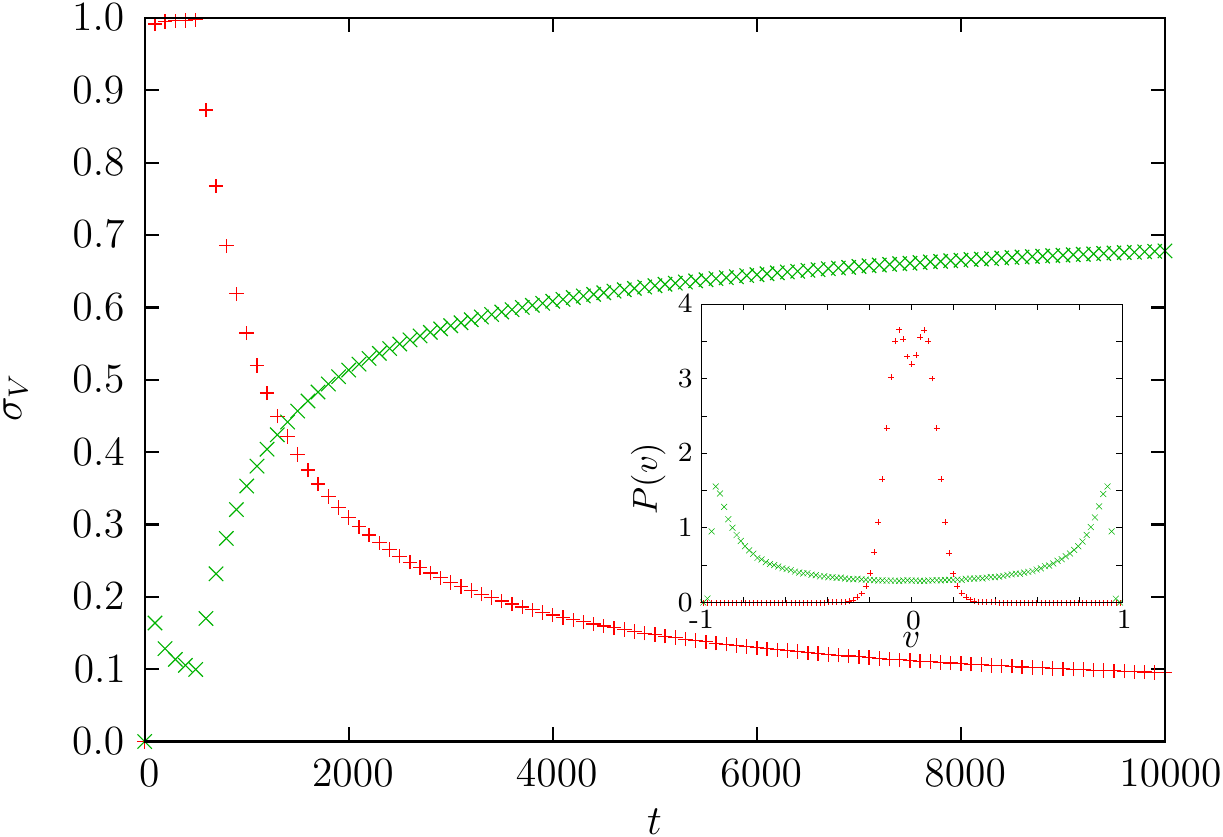}
\caption{
Same as figure~\ref{f:control} but for Gaussian utility distribution ($\mu=1$, $\varsigma=1$).   Note that in the low-to-high noise case with these parameters, two peaks are still visible in the velocity histogram at $t=10000$; they are a residual effect from the first 500 time steps and merge if the final time is further increased.
}
\label{f:controlG}
\end{figure}

A similar argument applies to other distributions with exponential tails -- the characteristic largest value of $U^\pm$ converges to the mode of the corresponding Gumbel distribution and generically grows as $(\ln X^\pm )^\kappa$ where the power $\kappa$ determines the long-time stability of the symmetric fixed point via
\begin{equation}
\Delta'_t(0) \sim \frac{(\ln t)^{\kappa-1}}{ T}.
\end{equation}
The exponential distribution~\eref{e:exp} corresponds to the special case of $\kappa=1$, while for $\kappa<1$ we expect long-term stability  of the symmetric fixed point ($\Delta'_t(0)<1$) and, for $\kappa > 1$ we expect long-term instability ($\Delta'_t(0)>1$).  For intermediate timescales, the system can be driven towards either the symmetric mixed state or the asymmetric frozen state by increasing or decreasing the noise, as we demonstrate for the Gaussian distribution in figure~\ref{f:controlG}.

\subsection{Power-law tails}
\label{ss:pow}

The second class of extreme value statistics corresponds to distributions with power-law tails as typified by the Pareto distribution with c.d.f.\
\begin{equation}
F(u) = 
1 - \left( \frac{u_m}{u} \right)^\alpha, \quad u \geq u_m 
\end{equation}
where $u_m$ is a lower bound and $\alpha > 0$.  In this case one finds that the characteristic largest value after $X^\pm$ steps is given by $u_m (X^\pm)^{1/\alpha}$ leading to the approximation
\begin{equation}
P^\pm = \frac{\rme^{u_m (V^\pm t)^{1/\alpha}}}{\rme^{u_m (V^+ t)^{1/\alpha}} + \rme^{u_m (V^- t)^{1/\alpha}}}
\end{equation}
and hence, by the same method as previously, 
\begin{equation}
\Delta_t(v)=\tanh \left( \frac{u_m t^{1/\alpha}}{2 T} \left\{ \left[\frac{1+v}{2} \right] ^{1/\alpha} - \left[\frac{1-v}{2}\right]^{1/\alpha} \right\} \right).
\end{equation}
Again, for all $t$ we find a symmetric fixed point at $v^*=0$, here with stability determined by the slope
\begin{equation}
\Delta'_t(0) = \frac{u_m}{T \alpha} \left(\frac{t}{2}\right)^{1/\alpha}
\end{equation}
which is greater than unity for $t > 2  (T \alpha / u_m)^\alpha $.  In fact, in the limit $t \to \infty$, $\Delta_t(v)$ approaches the step function $\textrm{sgn}(v)$ with corresponding stable fixed points at $v^* = \pm 1$.  

\begin{figure}
\centering
\includegraphics[width=0.8\textwidth]{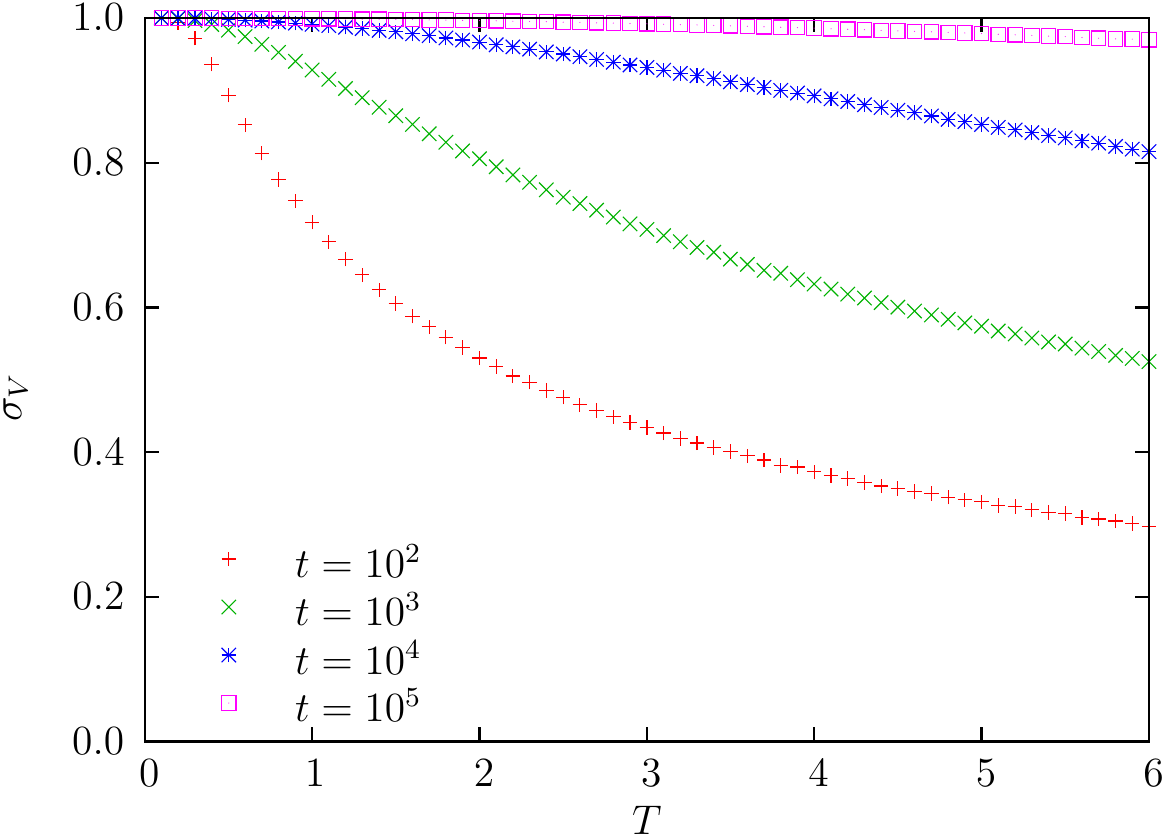}
\caption{
Same as figure~\ref{f:range} but for Pareto utility distribution ($x_m=0.5$, $\alpha=2$).
}
\label{f:rangep}
\end{figure}

In this Pareto case, it is straightforward to show that for large $X^\pm$ the maximum value $U^\pm$ has approximately a Fr\'echet c.d.f.\
\begin{equation}
G(u^\pm) = \rme^{-\left( u^\pm/ s^\pm \right)^{-\alpha}} \quad u^\pm > 0 
\end{equation}
where the scale parameter $s^\pm$ is given by $u_m (X^\pm)^{1/\alpha}$.  The mean of this distribution is only finite for $\alpha > 1$ but the mode and the median are both proportional to $s^\pm$ so, again, the trivially calculated characteristic largest value should give a good indication of the long-time behaviour.   This is confirmed in figure~\ref{f:rangep} where the standard deviation of the velocity against noise strength is plotted for a case where the utility has a Pareto distribution with unit mean ($x_m=0.5$, $\alpha=2$).  
For all values of $T$, the velocity variance converges towards unity (corresponding to individual trajectories approaching the asymmetric fixed points at $v^*=\pm 1$).  We have also checked that the convergence is faster for smaller values of $\alpha$ (``longer tails''), noting in particular that the distribution of $U$ has infinite mean for $\alpha \leq 1$.

More generally, the Fr\'echet distribution is the limiting form for the rescaled maximum of i.i.d.\ random variables drawn from any distribution with power-law tails~\cite{Fisher28,Gnedenko43}.   In all such cases we expect that $\Delta'_t(0)$ increases as some power of $t$, leading each agent to ultimately become frozen in a pure state corresponding to one or other choice.  We remark that this power-law dependence  is stronger than the logarithmic form found in section~\ref{ss:exp}; even by increasing the noise we only expect to be able to favour the mixed state for short timescales, e.g., up to the order of $(T \alpha / u_m)^\alpha$ for the Pareto distribution considered above.

\subsection{Bounded distributions} 

Finally, we consider distributions of $U$ with finite upper bound (as might be appropriate, for instance, if an agent's memory is based on some predetermined numerical scale with given minimum and maximum).   The obvious example is a uniform distribution with c.d.f.\
\begin{equation}
F(u) = \cases{ \case{u-l}{r-l} & $ u \in [l,r)$ \\
 1 & $u \geq r$ \\}
\end{equation}
whose characteristic largest value after $X^\pm$ steps is given by $r - (r-l)/X^\pm$.  Notice that, in contrast to the previous examples, this converges to a finite constant as $X^\pm \to \infty$ which is an elementary consequence of the upper bound on the underlying distribution and already gives a hint at the long-time behaviour.  

\begin{figure}
\centering
\includegraphics[width=0.8\textwidth]{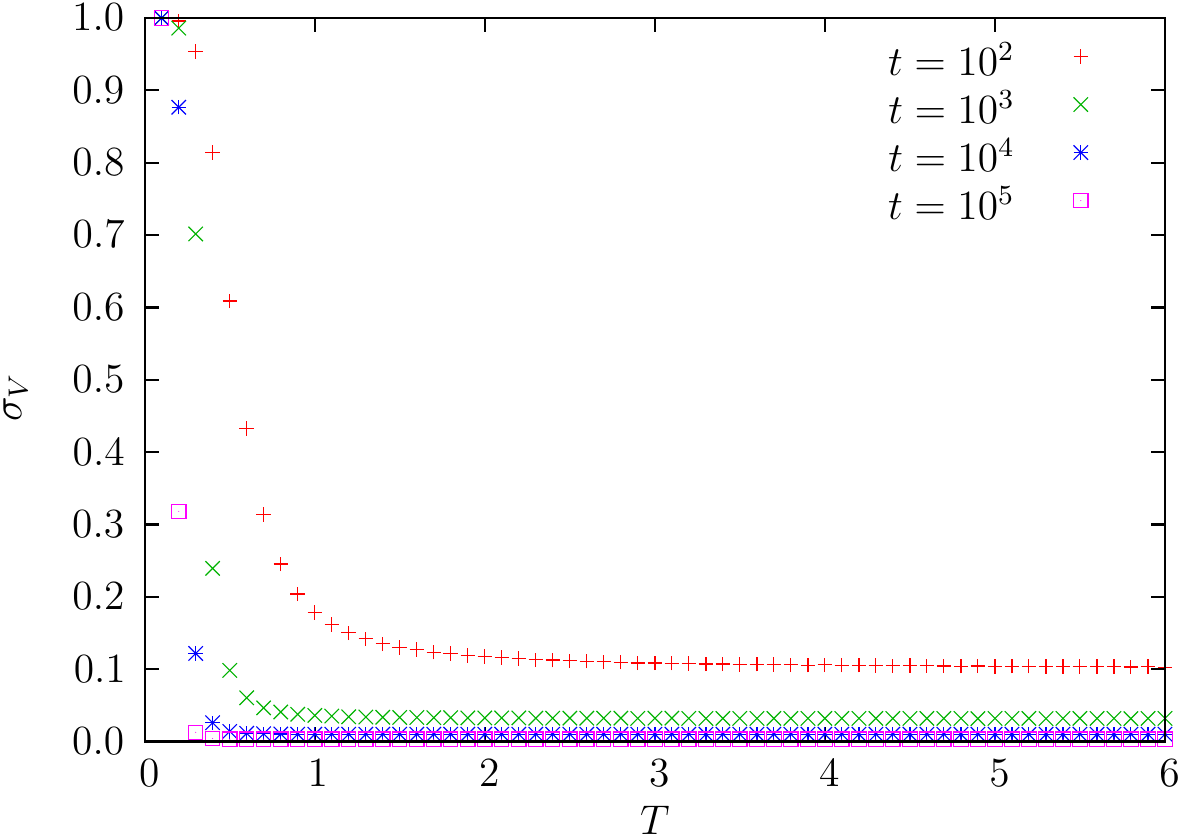}
\caption{
Same as figure~\ref{f:range} but for uniform utility distribution ($l=0$, $r=2$).
}
\label{f:rangeu}
\end{figure}

\begin{figure}
\centering
\includegraphics[width=0.8\textwidth]{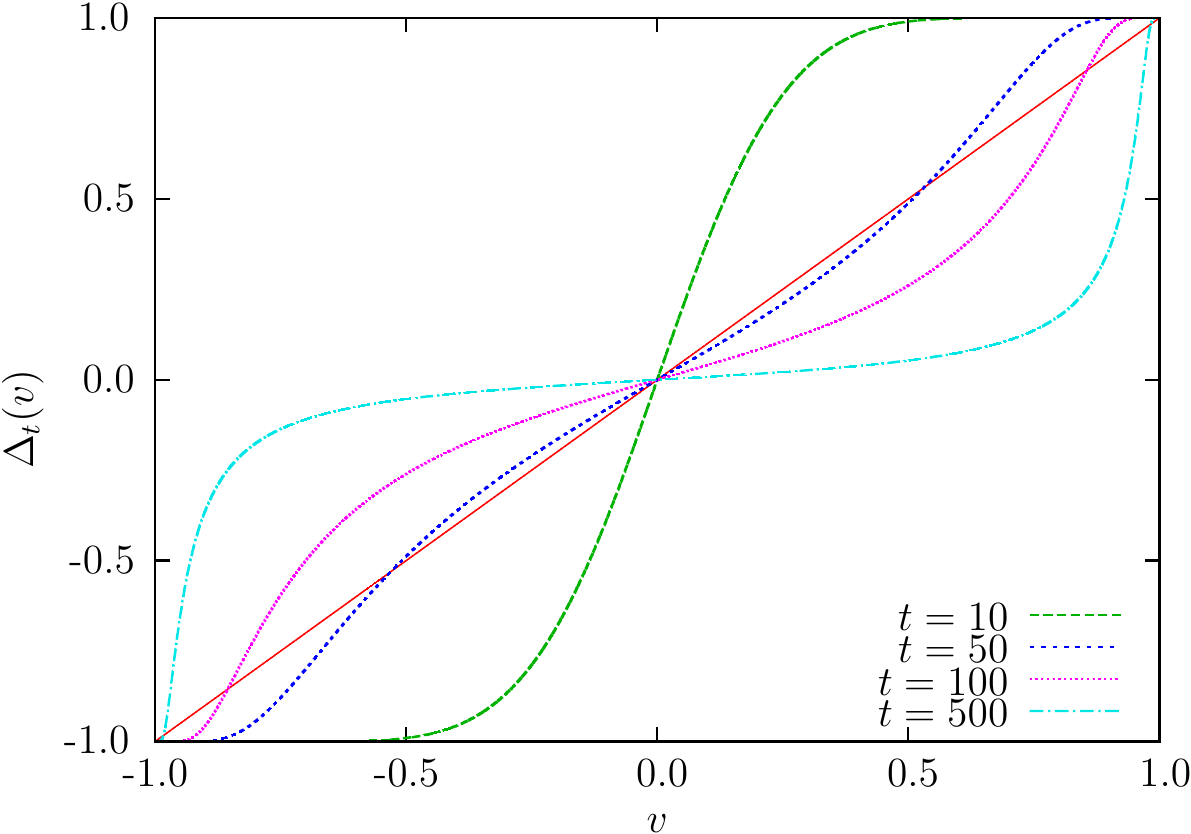}
\caption{Plot of $\Delta_t(v)$ given by~\eref{e:deltaun} for low noise and increasing times.  Note intersections with diagonal $v$ (solid red line) and compare sketch graphs in figure~\ref{f:cobweb}.  Parameters are $l=0$, $r=2$, and $T=0.1$.}
\label{f:uncob}
\end{figure}

In this case, following our previous heuristic procedure we have
\begin{equation}
P^\pm = \frac{\rme^{-(r-l)/ (T V^\pm t)}}{\rme^{-(r-l)/ (T V^+ t)}+\rme^{-(r-l)/ (T V^- t)}}
\end{equation}
and
\begin{equation}
\Delta_t(v)=\tanh \left( \frac{r-l}{ T t} \left\{ \frac{1}{1-v} - \frac{1}{1+v}  \right\} \right). \label{e:deltaun}
\end{equation}
The slope at the symmetric fixed point is given by 
\begin{equation}
\Delta'(0)=\frac{2(r-l)}{ T t}
\end{equation}
which is less than unity for $t > 2 (r-l) /  T $ and tends to zero as $t \to \infty$.   Hence we argue that the symmetric fixed point is always stable for long enough times (regardless of noise strength).  This conclusion is supported by the simulation data in figure~\ref{f:rangeu}.
The observed behaviour of the variance for very small $T$ can be roughly explained by noting that, for this version of the model, the walker can become stuck for finite times in a metastable fixed point at $v^*=\pm 1$.\footnote{In fact, initial effects may also play a complicating role here since the characteristic largest value argument is strictly only a good approximation when the number of steps in both directions is large.}  To see this, we plot in figure~\ref{f:uncob} the function $\Delta_t(v)$ of~\eref{e:deltaun} and examine its intersections with the line $v$, for fixed $T$ and increasing $t$.
Notice that, in this case, for $t > 2 (r-l) / T$ both symmetric \emph{and} asymmetric fixed points are stable but separated by an unstable point whose position tends to $\pm 1$ as $t \to \infty$.  The corresponding potential landscape has metastable states at $v^*=\pm 1$ and a trajectory can be trapped in such a state until fluctuations drive it over the barrier (whose relative height decreases with time) to the global minimum at $v^*=0$.  We emphasize that, since the fixed point at $v^*=0$ is always stable except for very short times, the long-time behaviour of the system cannot be effectively controlled by altering the noise (confirmed by further simulations, not shown).

It is easy to show that, for large $X^\pm$, the maximum of i.i.d.\ uniform random variables has approximately the reversed (unit) Weibull distribution
\begin{equation}
G(u^\pm) = \rme^{(u-r)/s^\pm} \quad u_\pm \leq r
\end{equation}
with scale parameter $s^\pm=(r-l)/X^\pm$ and mean coinciding with the characteristic largest value calculated above.
However, once again, the argument is more broadly applicable -- for bounded distributions the limiting distribution of the rescaled maximum is generically reversed Weibull (also known as ``Type III'' extreme value) with mean and median typically approaching the upper bound as some inverse power of the number of trials~\cite{Fisher28,Gnedenko43}.  In all such cases, $\Delta'_t(0) \to 0$ as $t \to \infty$, meaning each agent is expected to ultimately end up in the mixed state with both choices equally likely.  Physically, it is clear that in the long-time limit the system approaches a standard memoryless diffusive model with $P^+$ and $P^-$ fixed and equal.

\section{Discussion}
\label{s:dis}

\begin{figure}
\centering
\includegraphics[width=0.8\textwidth]{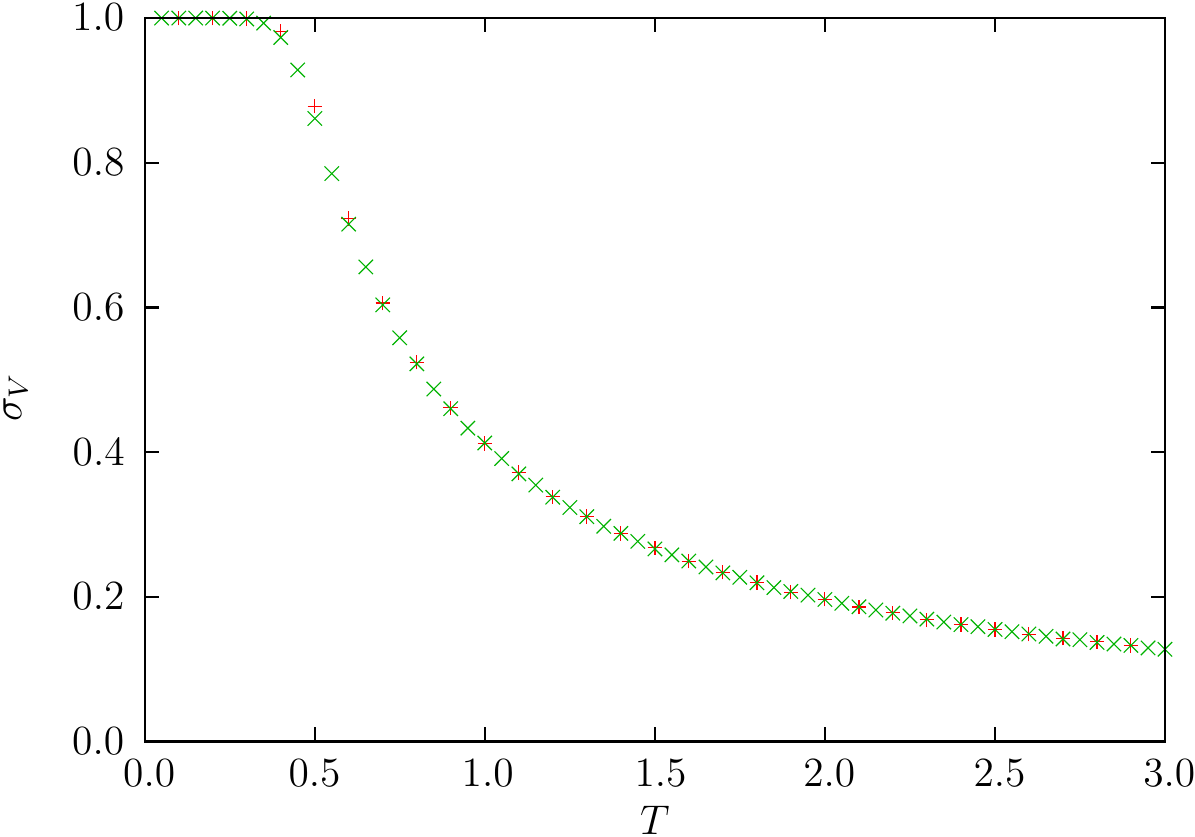}
\caption{Standard deviation of final velocity $V=V^+-V^-$ for random walkers with dependence on peak \emph{and} end values of exponential utility distribution ($\lambda=1$) and range of noise values.  Red points ($+$ symbols) show simulation data for the peak-end model while green points ($\times$ symbols) show comparative results for peak dependence only but with noise $2T$.  Both cases calculated from $10^5$ trajectories and $10^5$ time steps.}
\label{f:PE}
\end{figure}

In this paper we have performed a detailed analysis of random walkers with peak memory dependence.  Commensurate with the original motivation of Kahneman's peak-end rule, we now make some comments on the effect of also including an explicit dependence on the final value of the utility $U$.  To be precise, we consider the mean of peak and final experience so that the right and left hopping probabilities in~\eref{e:logit} are replaced by
\begin{equation}
\fl
P^+ = \frac{\rme^{(U^++U^+_f)/(2T)}}{\rme^{(U^++U^+_f)/(2T)}+\rme^{(U^-+U^-_f)/(2T)}}, \qquad P^- = \frac{\rme^{(U^-+U^-_f)/(2T)}}{\rme^{(U^++U^-_f)/(2T)}+\rme^{(U^-+U^-_f)/(2T)}} \label{e:logit2}  
\end{equation}
where here $U^+_f$ ($U^-_f$) is the value of $U$ corresponding to the last step right (left).  Note that the $U^\pm_f$ are in general smaller and much less strongly correlated than the $U^\pm$ so we might expect their effect to cancel out on average in the long-time limit.  At the same time, the dependence on $U^\pm$ is here weakened in the sense that the values are now divided by $2T$ rather than $T$.  The simulation results in figure~\ref{f:PE}, for the case of an exponential utility distribution, confirm that this modified model behaves very similarly to an increased noise version of the original model with the replacement of $T$ by $2T$.  

With the preceding paragraph in mind, we argue that our work on the peak memory model also has implications for the peak-end case.  Specifically, we have found that the effect of noise/disruption in the model is dependent on the properties of the utility distribution.   Using the characteristic largest value to cast the problem as an effective P\'olya process 
provides direct information on the long-time dynamics (in particular the stability of fixed points in the system) but, in order to quantify the observed variance, one also needs to consider the distribution of maximum values.  The examples we have shown, together with general arguments rooted in extreme value theory, reveal three qualitatively different classes of behaviour:
\begin{itemize}
\item For utility distributions with \textbf{heavy tails} each random walker (agent) eventually becomes frozen in a state corresponding to one or other step direction (choice), regardless of the level of noise.
\item For \textbf{bounded} utility distributions each agent samples both choices approximately equally in the long-time limit, again regardless of the level of noise.
\item For utility distributions with \textbf{exponential tails} the situation is more subtle -- for an $\rme^{- \lambda u}$ decay we find a transition between frozen and mixed states at $\lambda T=1$; in other cases there is a weak logarithmic dependence on the time.  Furthermore, for the special case of $\rme^{- \lambda u}$ decay, even in the high-noise regime there is a finite variance around the mixed state which can be attributed to fluctuations in the maximum values.
\end{itemize}
Significantly, this implies that only for exponential-tailed utility distributions can one hope to increase the switching between decisions on intermediate/long timescales simply by increasing the noise.

From a statistical physics point of view, it would be interesting in the exponential case to characterize the phase transition and scaling exponents at $\lambda T=1$, e.g., by calculating the correlation function.  This latter is also relevant in the opinion dynamics context as it quantifies how sensitive the long-time behaviour is to the first step and thus the extent to which a particular one of the two asymmetric fixed points might be favoured by a small initial perturbation.  Preliminary simulations suggest that the correlation function in the full model converges to zero for $\lambda T > 1$ and, for $\lambda T \ll 1$, decreases more strongly with $\lambda T$ than in the artificial model of~\ref{A:det} (presumably due to the added fluctuations reducing the effect of the initial conditions).    However, a more detailed analysis with finite-time scaling along the lines of~\cite{Mori15} is deferred to a future publication.

Other extensions of the work might include considering coupled random walkers (modelling collective rather than individual memory) or peak effects in other opinion dynamics models, such as contact processes and voter models~\cite{Liggett97}.    The peak-end rule itself can also be critiqued (see, e.g., the discussion in~\cite{Kemp08}) and realistic refinements such as the slow fading of peak memories in the distant past could be incorporated into the modelling.  However, we believe that our current work represents an important first step beyond simply averaging over the whole past experience or just recalling the most recent history.

Although our analytical calculations thus far have been carried out in the framework of a specific toy model they highlight more generally the possible role of (experienced or remembered) utility distributions in maintaining and controlling behaviour.  In particular, if agents are encouraged to reflect on their own experiences with a view to possibly modifying future choices then the outcome could be subtlely dependent on any numerical scale offered for reflection, e.g., whether or not it has a fixed upper bound.  There is much scope here for future interdisciplinary work linking with current understanding in psychology and economics.

\ack
This research was carried out as part of the project \emph{Reflect -- A feasibility study in experienced utility and travel behaviour} funded by Research Councils UK (EP/J004715/1).  The culmination of the work also benefited from the kind hospitality of the Galileo Galilei Institute for Theoretical Physics (GGI) Florence and the National Institute for Theoretical Physics (NITheP) Stellenbosch.  The author wishes to thank Jennifer Roberts and Hugo Touchette for many helpful discussions as well as comments on a draft manuscript.  Raw data underlying the figures shown here is stored in the Queen Mary Research Online repository (http://dx.doi.org/10.17636/01007419).

\appendix

\section{Artificial model for comparison}
\label{A:det}
Here we present a brief analysis of a simplified model in which $U^\pm$ at every time step is deterministically given by $({\ln X^\pm})/{\lambda}$ (the characteristic largest value of an exponential distribution with $X^\pm$ trials).  In this case, the future hopping rates are completely determined by the past velocity so previous work on generalized P\'olya urn models, etc.\ should be directly applicable. 

\begin{figure}
\centering
\includegraphics[width=0.8\textwidth]{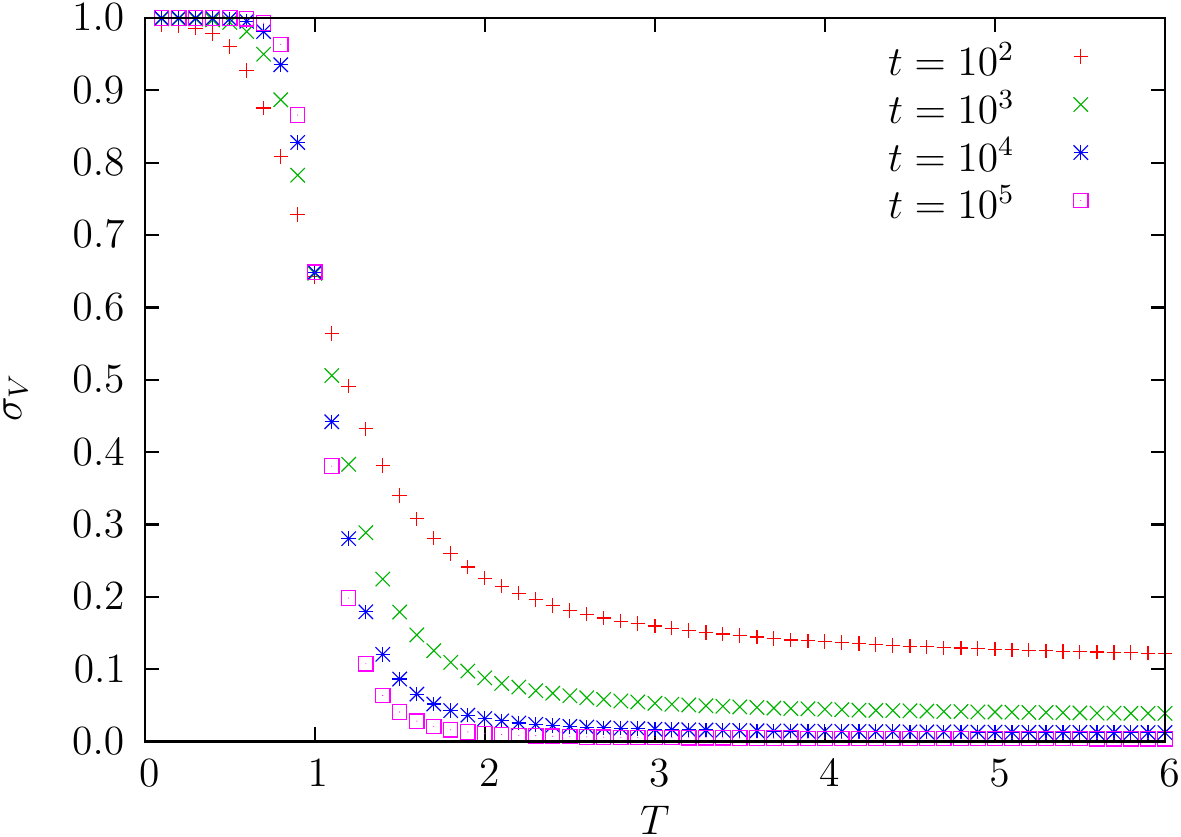}
\caption{
Same as figure~\ref{f:range} but for simplified model (still with $\lambda=1$).
}
\label{f:prange}
\end{figure}

\begin{figure}
\centering
\includegraphics[width=0.8\textwidth]{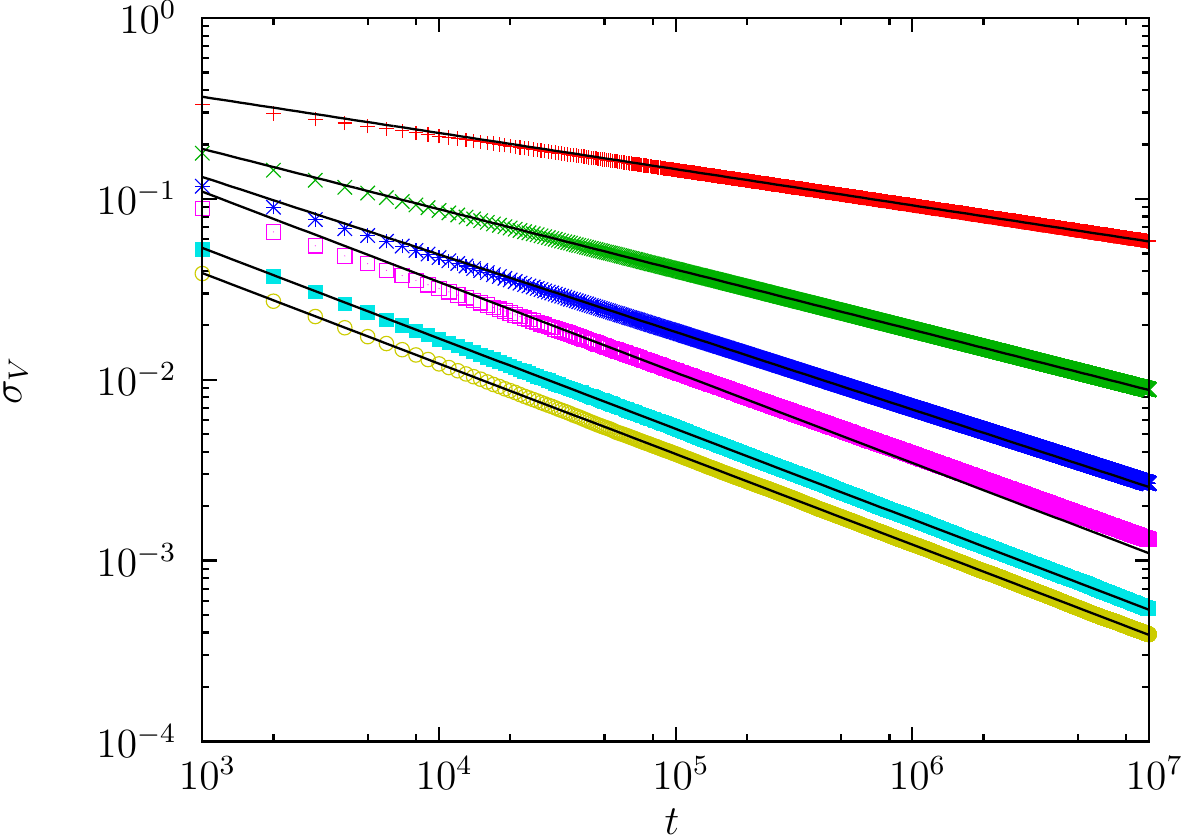}
\caption{Standard deviation of velocity $V=V^+-V^-$ against time for random walkers in simplified model ($\lambda=1$) at selected noise values ($10^4$ trajectories in each case).  Points are simulation data for (top to bottom): $T=1.25, 1.5, 1.75, 2, 3, 6$.  Black solid lines are fits corresponding to power laws with negative exponent $\min(1/2,1-1/(\lambda T))$; logarithmic corrections are expected at $\lambda T =2$.}
\label{f:decay}
\end{figure}

The standard deviation against noise in this case is shown in figure~\ref{f:prange} which is to be compared with figure~\ref{f:range} in the main text. 
For the artificial model, the standard deviation \emph{does} appear to converge to unity for $\lambda T < 1$ and zero for $\lambda T > 1$ as predicted by the analysis of typical trajectories outlined in section~\ref{ss:method}.  In figure~\ref{f:decay} we examine on log-log scale the limiting behaviour for selected points in the high-noise regime and clearly see a power-law decay.  In fact, the asymptotic behaviour of the variance can be predicted on general grounds to depend on the slope $\Delta'(v^*)$ at the stable fixed point (cf.~\cite{Me09,Me15unpub} for the continuous-time case).  Specifically, as found in other models~\cite{Schutz04,Hod04,Huillet08b}, one anticipates a dynamical phase transition at $\Delta'(v^*)=1/2$ with diffusive fluctuations (i.e., $t^{-1/2}$ decay of the velocity standard deviation) for $\Delta'(v^*)<1/2$ and superdiffusive behaviour (with $t^{\Delta'(v^*)-1}$ decay) for $\Delta'(v^*)>1/2$.   Setting $\Delta'(v^*)=1/(\lambda T)$, the resulting predictions indeed fit the simulation data very well (with logarithmic corrections expected at the dynamical phase transition itself).  Note that this provides a quantitative explanation for the observed slow convergence close to the transition point ($\lambda T = 1$) which may also be relevant in the full model.

Another important quantity for generalized P\'olya processes such as this is the correlation function $C(t)$ between the direction of the first step and the $(t+1)$th step~\cite{Mori15}.   The long-time limit of this quantity plays the role of an order parameter and simulation results in the present case (not shown) confirm a continuous phase transition at $\lambda T=1$ with $C(t)$ converging to zero for all $\lambda T>1$.

\section{Limiting variance in exponential case}
\label{A:var}

To understand the fluctuations of the velocity, one needs to take account of the fact that each value of $U^\pm$ persists for a number of time steps before being replaced by a larger one.  In this appendix, we pursue such an approach with an exponential utility distribution to obtain an approximation for the long-time limit of the standard deviation in the high-noise regime.

We let $\tau$ be the time at which the last record occurred (i.e., the last change in either $U^+$ or $U^-$ up to the current time $t$) and note that, by definition, the value of $P^+-P^-$ is unchanged for $\tau \leq i \leq t$.  In the exponential case  $U^+-U^-$ is given by~\eref{e:eps} and typical trajectories should then obey the \emph{stochastic} mapping
\begin{equation}
v_{t+1}=\frac{\tau v_\tau + (t+1-\tau) \tanh\left( \frac{1}{2\lambda T} \ln \left( \frac{1+v_t}{1-v_t}\right) + \frac{\epsilon}{2T} \right)}{t+1}
\end{equation}
where $\tau$ and $\epsilon$ are random variables and the second term in the numerator gives the expected displacement since the last record.  In the case where $\tau=t$ and $\epsilon=0$ for all time (i.e., $U^+$ or $U^-$ updated at every step with no fluctuations) we recover the deterministic mapping of~\eref{e:dismap} with~\eref{e:Delta}.  In slight abuse of notation, we denote a typical trajectory by $v_t$ even in the present stochastic case and argue that it is the fluctuations of this trajectory which lead to the long-time variance of $V$.

For large times we can approximate $t+1 \approx t$ and $v_{t+1} \approx v_t$ (time-averaged velocity changes slowly) which allows us to write
\begin{equation}
v_t \approx \rho_{\tau} v_\tau + (1-\rho_\tau) \tanh\left( \frac{1}{2\lambda T} \ln \left( \frac{1+v_t}{1-v_t}\right) + \frac{\epsilon}{2T} \right)
\end{equation}
with $\rho_\tau$ the fraction $\tau/t$.  To proceed further we then make three key assumptions:
\begin{enumerate}
\item \label{a:linear} $|v_t|$ and $|\epsilon|$ are sufficiently small that we can approximate the $\tanh$ term by a linear function.
\item \label{a:uniform} The $X^+$ right steps (and therefore also the $X^-$ left steps) are uniformly distributed throughout the trajectory.
\item \label{a:indep} $v_\tau$, $\epsilon$ and $\rho_\tau$ are mutually independent.
\end{enumerate}
All three of these assumptions are expected to fail as $\lambda T$ approaches 1 but they do facilitate analytical progress for $\lambda T \gg 1$.

First, we use assumption~(\ref{a:linear}) to make a linear expansion
\begin{equation}
v_t \approx \rho_\tau v_\tau + (1-\rho_\tau)\left(\frac{v_t}{\lambda T} + \frac{\epsilon}{2T} \right),
\end{equation}
and trivially rearrange to obtain
\begin{equation}
v_t \approx \frac{\lambda T \rho_\tau}{\lambda T - 1 + \rho_\tau} v_\tau + \frac{1-\rho_\tau}{\lambda T - 1 +\rho_\tau} \left( \frac{\epsilon'}{2} \right)
\end{equation}
where for convenience we have rescaled to $\epsilon'$ which has a standard logistic distribution with variance $\sigma^2_{\epsilon'}=\pi^2/3$.  To obtain the distribution of $\rho_\tau$, we first consider separately the fractional times $\rho_\tau^+$ and $\rho_\tau^-$ for the last records corresponding to right and left steps respectively.  Each of the $X^\pm$ previous steps is equally likely to have produced the maximum value so, with assumption~(\ref{a:uniform}), we have $\rho^\pm_\tau \sim \textrm{Uniform}(0,1)$ in the long-time limit.  Then, by straightforward calculation, $\rho_\tau = \max(\rho^+_\tau,\rho^-_\tau)$ is governed by a triangular distribution on $[0,1]$ with mode 1. 

Now, in the long-time limit, we expect the distribution of $v_\tau$ to be the same as that of $v_t$, both characterized by standard deviation $\sigma_V$.  Hence using the independence assumption~(\ref{a:indep}), together with the symmetry of trajectories around zero, we obtain
\begin{equation}
\sigma_V^2 \approx p(\lambda T) \sigma_V^2 + q(\lambda T) \frac{\sigma_{\epsilon'}^2}{4}
\end{equation}
where the functions $p(x)$ and $q(x)$ are given by expectations with respect to the distribution of $\rho_\tau$:
\begin{eqnarray}
\fl 
p(x)&=E\left[ \left( \frac{x \rho_\tau}{x-1+\rho_\tau} \right)^2  \right] &= x\left[-2+9x-6x^2+6x(x-1)^2\ln\left(\frac{x}{x-1}\right)\right], \\
\fl q(x)&=E\left[ \left( \frac{{1-\rho_\tau}}{x-1+\rho_\tau} \right)^2 \right] &= 1-6x+2x(3x-2)\ln\left(\frac{x}{x-1}\right).
\end{eqnarray}
Rearranging and substituting in for $\sigma_{\epsilon'}$ yields the final approximation
\begin{equation}
\sigma_V\approx \frac{\pi}{2\sqrt{3}} \sqrt{\frac{q(\lambda T)}{1- p(\lambda T)}}. \label{e:finapprox}
\end{equation}

The expression in~\eref{e:finapprox} clearly diverges at $\lambda T=1$ but we can estimate its range of applicability by considering assumption~(\ref{a:linear}).  Specifically, $x$ approximates $\tanh(x)$ to within 10\% for $|x| \lesssim 0.55$ so, since $v_t$ and $\epsilon$ are fairly strongly correlated, we require
\begin{equation}
\frac{\sigma_V}{\lambda T} + \frac{\sigma_{\epsilon'}}{2 \lambda T} \lesssim 0.55
\end{equation}
which is satisfied for $\lambda T \gtrsim 2.3$.  Indeed~\eref{e:finapprox} is seen to provide a reasonable approximation to the simulation data in this regime (cf.\ figure~\ref{f:range}).  The small remaining discrepancy is probably mainly due to the failure of assumption~(\ref{a:indep}); in particular, $v_\tau$ is not strictly independent of $\epsilon$.

\section*{References}
\bibliographystyle{iopart-num}

\begin{thebibliography}{10}
\expandafter\ifx\csname url\endcsname\relax
  \def\url#1{{\tt #1}}\fi
\expandafter\ifx\csname urlprefix\endcsname\relax\def\urlprefix{URL }\fi
\providecommand{\eprint}[2][]{\url{#2}}


\bibitem{Bachelier1900}
Bachelier L 1900 {\em Annales Scientifiques de l'\'Ecole Normale
  Sup\'erieure\/} S\'er.\ 3 {\bf 17} 21--86

\bibitem{Chowdhury00}
Chowdhury D, Santen L and Schadschneider A 2000 {\em Phys. Rep.\/} {\bf 329}
  199--329

\bibitem{Bressloff14}
Bressloff P~C 2014 {\em Stochastic Processes in Cell Biology\/} ({\em
  Interdisciplinary Applied Mathematics\/} vol~41) (Springer)

\bibitem{Castellano09b}
Castellano C, Fortunato S and Loreto V 2009 {\em Rev. Mod. Phys.\/} {\bf 81}
  591--646

\bibitem{Hanggi78}
H\"anggi P 1978 {\em Z. Phys. B\/} {\bf 31} 407--416

\bibitem{Volkov83}
Volkov V~S and Pokrovsky V~N 1983 {\em J. Math. Phys.\/} {\bf 24} 267--270

\bibitem{Metzler99}
Metzler R, Barkai E and Klafter J 1999 {\em Phys. Rev. Lett.\/} {\bf 82}
  3563--3567

\bibitem{Hirschberg09}
Hirschberg O, Mukamel D and Sch\"utz G~M 2009 {\em Phys. Rev. Lett.\/} {\bf
  103} 090602

\bibitem{Concannon14}
Concannon R~J and Blythe R~A 2014 {\em Phys. Rev. Lett.\/} {\bf 112} 050603

\bibitem{Khoromskaia14}
Khoromskaia D, Harris R~J and Grosskinsky S 2014 {\em J. Stat. Mech.\/}  P12013

\bibitem{Schutz04}
Sch\"utz G~M and Trimper S 2004 {\em Phys. Rev. E\/} {\bf 70} 045101

\bibitem{Cressoni07}
Cressoni J~C, da~Silva M~A~A and Viswanathan G~M 2007 {\em Phys. Rev. Lett.\/}
  {\bf 98} 070603

\bibitem{Kenkre07}
Kenkre V~M 2007 Analytic formulation, exact solutions, and generalizations of
  the elephant and the {A}lzheimer random walks (\textit{Preprint}
  \eprint{arXiv:0708.0034})

\bibitem{Serva13b}
Serva M 2013 {\em Phys. Rev. E\/} {\bf 88} 052141

\bibitem{Polya31}
P\'olya G 1930 {\em Ann. Inst. Henri Poincar\'e\/} {\bf 1} 117–--161

\bibitem{Pemantle07}
Pemantle R 2007 {\em Probab. Surv.\/} {\bf 4} 1--79

\bibitem{Benjamini03}
Benjamini I and Wilson D~B 2003 {\em Electron. Commun. Probab.\/} {\bf 8}
  86--92

\bibitem{Basdevant08}
Basdevant A~L and Singh A 2008 {\em Probab. Theory Related Fields\/} {\bf 141}
  625--645

\bibitem{Veto08}
Vet\H{o} B and T\'oth B 2008 {\em Electron. J. Probab.\/} {\bf 13} 1909--1926

\bibitem{Erschler12}
Erschler A, T\'oth B and Werner W 2012 {\em Probab. Theory Related Fields\/}
  {\bf 154} 149--163

\bibitem{Nandori11}
N\'andori P 2011 {\em Probab. Theory Related Fields\/} {\bf 150} 373--403

\bibitem{Kahneman93}
Kahneman D, Fredrickson B~L, Schreiber C~A and Redelmeier D~A 1993 {\em
  Psychological Science\/} {\bf 4} 401--405

\bibitem{Fredrickson08}
Fredrickson B~L and Kahneman D 1993 {\em Journal of Personality and Social
  Psychology\/} {\bf 65} 45--55

\bibitem{Redelmeier96}
Redelmeier D~A and Kahneman D 1996 {\em Pain\/} {\bf 66} 3--8

\bibitem{Do08}
Do A~M and Wolford A~V~R~G 2008 {\em Psychonomic Bulletin and Review\/} {\bf
  15} 96--98

\bibitem{Kemp08}
Kemp S, Burt C~D~B and Furneaux L 2008 {\em Memory and Cognition\/} {\bf 36}
  132--138

\bibitem{Marsden13}
Marsden G and Docherty I 2013 {\em Transportation Research Part A: Policy and
  Practice\/} {\bf 51} 46--55

\bibitem{Glauber63}
Glauber R~J 1963 {\em J. Math. Phys.\/} {\bf 4} 294--307

\bibitem{Hill80}
Hill B~M, Lane D and Sudderth W 1980 {\em Ann. Probab.\/} {\bf 8} 214--226

\bibitem{Mori15}
Mori S and Hisakado M 2015 Correlation function for generalized {P}{\'o}lya
  urns: {F}inite-size scaling analysis (\textit{Preprint}
  \eprint{arXiv:1501.00764})

\bibitem{Gumbel04}
Gumbel E~J 2004 {\em Statistics of Extremes\/} reprint ed (Dover)

\bibitem{Me09}
Harris R~J and Touchette H 2009 {\em J. Phys. A: Math. Theor.\/} {\bf 42}
  342001

\bibitem{Touchette09b}
Touchette H 2009 {\em Phys. Rep.\/} {\bf 478} 1--69

\bibitem{deHaan06}
de~Haan L and Ferreira A 2006 {\em Extreme Value Theory -- An Introduction\/}
  Springer Series in Operations Research and Financial Engineering (Springer)

\bibitem{Fisher28}
Fisher R~A and Tippett L~H~C 1928 {\em Proc. Cambridge Phil. Soc.\/} {\bf 24}
  180--190

\bibitem{Gnedenko43}
Gnedenko B 1943 {\em Ann. Math.\/} {\bf 44} 423--453



\bibitem{Liggett97}
Liggett T~M 1997 {\em Ann. Probab.\/} {\bf 25} 1--29

\bibitem{Me15unpub}
Harris R~J In preparation

\bibitem{Hod04}
Hod S and Keshet U 2004 {\em Phys. Rev. E\/} {\bf 70} 015104

\bibitem{Huillet08b}
Huillet T 2008 {\em J. Phys. A\/} {\bf 41} 505005

\end{thebibliography}

\providecommand{\newblock}{}

\end{document}